\documentclass[review]{elsarticle}

\usepackage{lineno,hyperref}
\modulolinenumbers[5]
\usepackage[margin=1.0in]{geometry}
\usepackage{comment}
\usepackage{soul}
\usepackage{wrapfig}

\usepackage[cmex10]{amsmath}
\usepackage{amsfonts}
\usepackage{amssymb}
\usepackage{mathabx}
\usepackage{cases}
\usepackage{cancel}

\usepackage{caption}
\usepackage{subcaption}

\allowdisplaybreaks

\usepackage{amsthm}

\theoremstyle{remark}

\usepackage{amsmath}

\newcommand{\dsum}{\displaystyle\sum}

\usepackage{xcolor}
\usepackage{caption}
\usepackage{subcaption}

\usepackage{algorithm}

\usepackage{tabularx, booktabs}
\newcolumntype{L}{>{\raggedright\arraybackslash}X}
\newcolumntype{b}{X}
\newcolumntype{s}{>{\hsize=.5\hsize}X}
\usepackage{multirow}

\usepackage{enumitem}
\usepackage{nomencl}

\usepackage{algorithm,setspace}
\usepackage{algpseudocode}

\makenomenclature

\usepackage{etoolbox}
\renewcommand\nomgroup[1]{%
  \item[\bfseries
  \ifstrequal{#1}{B}{Parameters}{%
  \ifstrequal{#1}{C}{Variables}{%
  {%
  \ifstrequal{#1}{A}{Sets}{}}}}%
]}

\begin{document}

\begin{frontmatter}

\title{Estimating Distribution Grid Upgrade Costs based on County-level Building Electrification Projections}

\author{Miguel Heleno\fnref{lbnl}\corref{correspondingauthor}} 
\ead{miguelheleno@lbl.gov}
\author{Albane Serès\fnref{}}

\address[lbnl]{Lawrence Berkeley National Laboratory, Berkeley, CA 94720, USA.}

\cortext[correspondingauthor]{Corresponding author}

\begin{abstract}
Building electrification projections are typically developed at aggregated geographic levels to support bulk power system analysis, even though a significant portion of the infrastructure investments required to accommodate electrification occurs within distribution systems. Existing methodologies to disaggregate load growth projections onto distribution networks are based on context-specific analyses that require detailed data on building electricity consumption, distribution feeders, and geographic parcels, which are often proprietary and unavailable for policy and economic impact studies. This paper presents a methodology to disaggregate county-level building electrification projections onto representative distribution feeders and evaluate their impact on distribution upgrade costs. The proposed approach combines bulk system end-use electrification projections with representative feeder sets to generate building-level electrification measures. These measures are then used to create load profiles that are consistent with county-level projections while preserving the spatial diversity of electrification demand required for distribution infrastructure analysis. Using a case study of Genesee County, in Michigan, the results show that the proposed methodology is able to capture a diversity of feeder-level responses to electrification and represent the different realities that building electrification produces within the distribution system. This allows to represent localized load variations and capture more realistic distribution system upgrade costs driven by electrification.

\end{abstract}


\end{frontmatter}

\section{Introduction}

Distribution grid upgrades are expected to account for an increasing share of electricity sector investments as electrification expands across transportation, buildings, and industry. Recent studies project the United States electricity demand to increase by 20-40\% by 2050 \cite{AEO2025, Gagnon2024StandardScenarios}. Electrification affects not only the total amount of electricity consumed, but also the timing and location of demand, introducing new loading conditions and creating potential stress on distribution networks \cite{Li2024ImpactCongestion, Deetjen2021USImpacts}. Distribution infrastructure already represents the largest share of investor-owned utility capital expenditures in the U.S. \cite{EEI2019}, and these costs may increase further with the electrification of residential and commercial building end-uses \cite{Mascherbauer2025}, with significant impacts on the electricity rates of these consumer classes \cite{Sandoval2025}. Anticipating the distribution grid costs associated with building electrification is therefore necessary for evaluating pathways, informing policy decisions, and designing strategies to maintain affordability under these scenarios.

Existing approaches to estimate distribution upgrade costs are primarily based on historical utility expenditures. These methods use historical investment data to derive marginal distribution costs, typically expressed in \$/kW, and project future costs from forecasted increases in peak demand \cite{Fares2017TrendsUtilities, Li2024ImpactCongestion}. The underlying data are commonly obtained from regulatory filings such as FERC Form 1 \cite{OEDI:1994-2019}, filed by investor-owned utilities in the U.S.. This approach implicitly assumes that future utility investments will follow patterns similar to those observed historically and that distribution costs scale proportionally with aggregate increases in peak demand. This assumption introduces inaccuracies because, unlike transmission and bulk power system costs, distribution grid infrastructure costs are inherently locational. Distribution investments can be triggered by changes in coincidental and non-coincidental peak demand from relatively small groups of consumers that create stress in specific areas of the grid. This makes past marginal distribution values, expressed in \$/kW, a poor way of extrapolating future upgrade needs. For example, in a distribution grid with 1{,}000 consumers, a 5\% increase in peak demand can result either from 500 consumers increasing their consumption by 10\%, or from 50 consumers doubling their consumption at the peak time. Both cases produce the same aggregate 5\% peak increase, but the second case concentrates load growth in small locations and is more likely to trigger local distribution upgrades.

Recent works have adopted an alternative approach for estimating distribution upgrade costs by explicitly modeling future load growth on distribution feeders and calculating the resulting infrastructure upgrades. Several studies have used this methodology to evaluate the impacts of distributed photovoltaics \cite{dnv_2017}, electric vehicles \cite{2023KevalaCPUC}, and end-use electrification \cite{GUPTA2021116504, Palmintier2021}. These studies use feeder-level engineering models to estimate the specific equipment upgrades required under future load scenarios, closely resembling the distribution planning processes used by utilities \cite{pge_gna_2023}. Although this approach can provide detailed infrastructure costs, it requires feeder models and customer-level demand data that are often proprietary and difficult to obtain. One option to overcome the need for real feeders is to generate synthetic feeders from geographic information \cite{Mascherbauer2025, OEDI:AUS}, particularly when studying relatively small areas such as neighborhoods or cities. Another option is to use representative feeders calibrated for distribution grid economic and policy studies \cite{seres2025}, which can be systematically applied to larger geographic units such as counties or regions.

However, using synthetic or representative feeders in distribution impact studies requires disaggregating electrification projections, developed at the bulk power system level \cite{Langevin2026}, to the individual nodes of those feeders. This type of disaggregation is commonly performed within utility planning processes, where utilities develop electrification and load growth projections based on historical consumption, demographic and economic indicators, weather trends, and technology adoption assumptions \cite{pge_gna_2023, pacificorp_dsp_2021}. These projections are then disaggregated to the distribution system level using smaller geographic units and specialized tools such as LoadSEER \cite{hawaiianelectric_dsp_2021, pge_gna_2023}, where load forecasts are allocated to individual service transformers. These tools typically rely on methodologies based on statistical spatial regressions \cite{Bernards} or bottom-up agent-based models \cite{ROBINSON2015273} to generate future scenarios of load growth and technology adoption \cite{Sun2021, HELENO2020115826} or diffusion \cite{Ye2019, HEYMANN2019}. These methodologies are often context- and geography-specific and require detailed data on individual consumers, buildings, parcels, and network assets. As a result, they cannot be readily applied to systematic studies that rely on synthetic or representative feeder models of counties or regions.

To address this gap, this paper presents a methodology to disaggregate county-level building electrification projections onto representative feeders and evaluate their impact on distribution upgrade costs. The proposed methodology takes as input a set of representative feeders and end-use electrification projections and transforms them into individual building-level load profiles connected to feeder nodes. The resulting building-level profiles are consistent with the projected end-use energy consumption of the bulk power system electrification scenario while preserving the spatial diversity of electrification adoption across the feeder set. This creates localized changes in load that trigger representative feeder upgrades, which are then used to estimate distribution infrastructure costs. The specific contributions of this work are as follows:

\begin{itemize}
\item An optimization model to generate bottom-up building electrification measures at the distribution feeder level that minimize errors relative to aggregated county-level end-use electrification projections.

\item A simulation framework that takes building-level electrification profiles, assesses distribution system upgrade needs, and converts them into county-level distribution infrastructure costs.

\item A case study of Genesee County, Michigan that illustrates the methodology, demonstrates the diversity of feeder-level responses to electrification, and highlights the value of the proposed approach for estimating distribution grid upgrade costs.
\end{itemize}

\section{Proposed Methodology}
\label{sec: methods}

Estimating distribution grid capital investment costs due to building electrification requires representing building loads geographically in relation to distribution grid infrastructure, with sufficient granularity to produce realistic cost estimates. Estimating these costs at the county level requires generating these geographical representations consistently for county-level distribution grids and building stocks. To achieve this, we propose a three-step methodology:

\begin{itemize}
\item For each county, we create a set of representative distribution feeders, with building loads connected to their nodes. These feeders reflect the load characteristics of the county building stock in the baseline case and the distribution investment patterns historically observed in the corresponding climate zone.
\item Second, we disaggregate county-level high-electrification scenarios for each end-use onto the representative feeders by modifying the building load profiles and creating variations of individual building adoptions that match county-level projection. 
\item Third, we assess distribution grid needs for high-electrification scenarios using feeder-level analysis and produce distribution grid infrastructure cost estimates for the whole county.
\end{itemize}

\begin{figure}[ht]
    \centering
    \includegraphics[width=\linewidth]{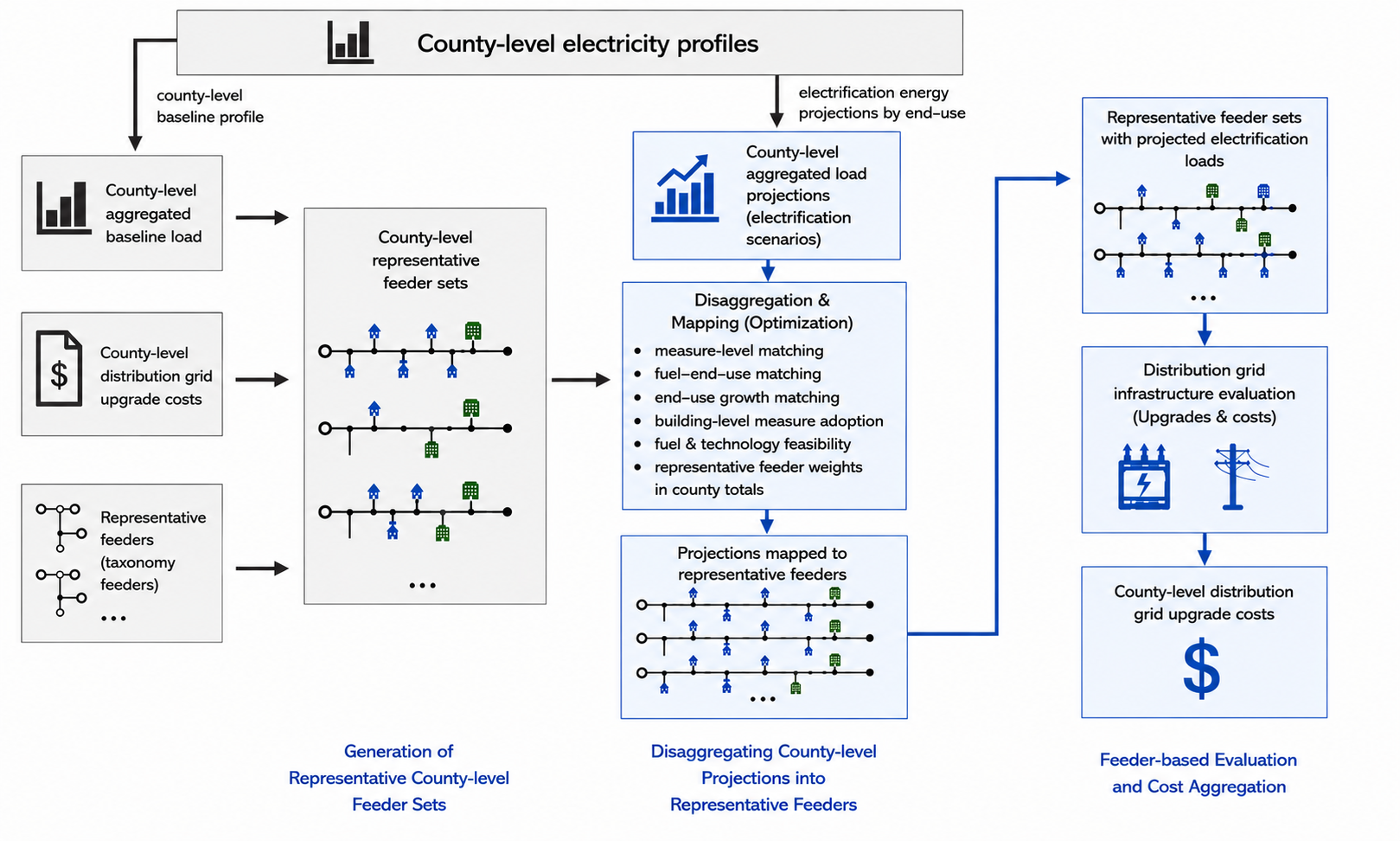}
    \caption{Methodology Overview.}
    \label{fig:methodology}
\end{figure}

Figure \ref{fig:methodology} provides an overview of the proposed methodology. The boxes depicted in grey illustrate the creation of representative county-level feeder sets, with nodal loads representing the county building stock, constructed to be consistent with the baseline of the county's electricity profile. This part of the methodology leverages our previous work \cite{seres2025} and aims to represent the baseline distribution system conditions of the county through representative feeder sets. The second part of the methodology, depicted in blue,  summarizes the contributions introduced in this work. Here, we translate the aggregated electrification projections into building-level electrification measures that can be assigned to individual buildings and feeders. This disaggregation preserves consistency between the baseline representative feeder sets and the projected county-level electrification scenario. Finally, the resulting feeder sets, updated with electrified building loads, are then used to assess distribution system upgrade needs and estimate distribution infrastructure costs.

\subsection{Generation of county-level representative feeders}
\label{section:methodology_feeder_sets}

To generate county-level representative feeders, we rely on the methodology presented in our previous work \cite{seres2025}. This methodology generates sets of representative feeders in the contiguous U.S., with county-level granularity, for distribution economic and policy studies. For a given county, it produces a population of representative feeders that captures the demand characteristics of the county (including energy, peak, and building types) and reproduces the economics and distribution infrastructure investment patterns historically observed in the area. Here, we provide a summary of this methodology and refer the reader to \cite{seres2025} for more details.

To generate feeder sets that reproduce the economics of local distribution grids, we start from the U.S. taxonomy feeders \cite{Schneider2008ModernReport}, which include 24 prototypical radial distribution feeders designed to represent typical rural and suburban systems across the five climate zones defined in \cite{Briggs2003ClimateClassification}. These feeders are categorized into three voltage classes (12.47 kV, 25.00 kV, 35.00 kV) and incorporate climate-driven heating and cooling demand patterns that influence feeder design. For large counties with significant urban populations, we also include the IEEE 342-Low Voltage Network Test System (LVNTS) \cite{lvnts}, a highly meshed, unbalanced 120/208 V urban network with eight 13.2 kV primary feeders supplying a 50 MVA load, representative of dense downtown areas.

Keeping the original topology of the taxonomy feeders, we generate random variations of each prototypical feeder by modifying its electrical equipment ratings so that the investments required under peak load growth match historically observed utility investment rates in each climate zone. To do this, we apply an artificial 10\% increase in feeder load and measure the required distribution grid upgrades using the methodology in \cite{seres2025} and the equipment reference costs in \ref{app:unit_costs}. If the resulting \$/kW rate is lower than the historical benchmark, we randomly decrease the capacity of selected conductors or transformers to increase upgrade needs under load growth. Conversely, if the resulting \$/kW rate is higher than the benchmark, we increase selected conductor or transformer ratings and add capacitors to reduce upgrade requirements. To obtain reference rates, we compile historical peak load and distribution grid upgrade costs from investor-owned utilities and estimate the investment rate associated with peak growth using the methodology and FERC data presented in \cite{OEDI:1994-2019}.

Using these climate zone feeders with calibrated equipment, we then allocate county-specific residential and commercial building profiles to feeder nodes, respecting the original sector load configurations specified in \cite{Schneider2008ModernReport}. We randomly sample county-specific ResStock~\cite{White2024ResStockDocumentation} and ComStock~\cite{osti_2522809} building profiles and assign them to nodes of each feeder by minimizing the nodal power assignment error. By repeating this process multiple times, we create a large pool of randomly-generated taxonomy feeders with county-representative buildings assigned to their nodes. Then, as described in \cite{seres2025}, we apply an optimization model to attribute a weight to each feeder in the pool so that the final resulting feeder weighted set matches the total energy and peak demand of the county.

\subsection{Disaggregating county-level electrification projections onto representative feeders}

We consider aggregated annual electricity consumption projections by end-use for the residential and commercial sectors that result from the electrification scenario for a given county, such as the ones presented in \cite{Langevin2026}. We also consider a set of electrification measures, each associated with a technology efficiency improvement or fuel-switching upgrade, that are available for each type of building in the building stock. In our work, we use the measures available in ResStock and ComStock for residential and commercial buildings, respectively, but the underlying proposed methodology can use any set of applicable measures. To evaluate the distribution grid impacts of these projections using the county-specific representative feeder sets described in Section~\ref{section:methodology_feeder_sets}, it is necessary to disaggregate the projected end-use consumption to the individual buildings represented in the feeder set. To achieve this disaggregation, we identify which buildings in which feeders adopt each electrification measure such that the total electricity consumption across all buildings in the representative feeder set matches the county-level electricity projections for the high-electrification scenario.

We start by organizing these county-level projections \cite{Langevin2026} per end-use and matching them with end-uses in ResStock and ComStock. We consider electrification measures associated with the following residential end-uses: HVAC (including cooling and heating subcategories of end-use), water heating, cooking, and dryers. In the commercial sector, we only consider commercial HVAC, including cooling, heating, and ventilation categories. The remaining end-uses listed in the electrification projections, such as refrigeration, computers and electronics, lighting, or cooking, do not have technologies that can be upgraded in the ResStock and ComStock datasets used in this work; therefore, they are treated as “Others.”

We then match buildings in the county-level representative feeder set with candidate electrification measures that are applicable to each building’s end-use and fuel type. We use the ResStock or ComStock metadata to capture each building's end-use electricity consumption if a given electrification measure is applied ($e^{elec}_{m,f,eu,b}$). We always include in the set of measures the reference case ($'Ref'$) of no adoption, in which end-use consumption remains constant in the electrification scenario.

Thus, for each county, represented by the feeder sets generated section \ref{section:methodology_feeder_sets}, we obtain the following data per residential and commercial end-use:

\begin{itemize}
    \item A list of possible electrification measures. Each measure ($m$) corresponds to a technology efficiency or fuel-switching upgrade, that can be applied to an existing fuel ($f$) in a building of the county representative feeder sets ($b$). These measures will modify the electricity consumption of an end-use ($eu$). Measures associated to the HVAC end-use impact both heating and cooling subcategories.
    \item The electric energy consumption of the end-use $eu$ of each building, if  given measure $m$ is applied to convert the existing fuel $f$ currently serving the end-use ($e^{elec}_{m,f,eu,b}$).
    \item The total county-level projected electricity consumption associated with each end-use, if possible broken down by measure and fuel, in the high-electrification scenario ($E^{elec}_{m,f,eu}$) and in the base year ($E^{base}_{m,f,eu}$) used as a reference for the projections. If these county-level energy values are not provided by measure and fuel, they can be aggregated by end-use only.
    \item The total county-level electricity consumption per end-use category that is represented in the feeder sets, calculated in section \ref{section:methodology_feeder_sets} ($\hat{E}_{eu}$).
 
\end{itemize}

With these elements, we can disaggregate the county-level projections of the electrification scenario to individual buildings ($b_d$) in each feeder ($d$) of the representative set. To achieve this, we formulate an optimization model that selects the combination of measures, applied to each building end-use, that minimizes the difference from the aggregated projections in the high-electrification scenario.

{\setlength{\jot}{8pt}
\begin{align}
    & \min  \quad \beta_A \dsum_{e \in EU}\dsum_{f \in F}\dsum_{m \in M}\epsilon^A_{m,f,eu} + \beta_B \dsum_{e \in EU}\dsum_{f \in F} \epsilon^B_{f,eu} + \beta_C \dsum_{e \in EU} \epsilon^C_{eu} \label{eq:obj}
    \\
    & \left| \frac{\dsum_{d \in D}\dsum_{b \in B_d}{x_{m,f, b, d} \cdot w_d \cdot e^{elec}_{m,f,eu,b}} - E^{elec}_{m,f,eu}}{E^{elec}_{m,f,eu}} \right| \leq \epsilon^A_{m,f,eu} \label{eq:bdg} 
    \\
    & \left| \frac{\dsum_{m \in M}\dsum_{d \in D}\dsum_{b \in B_d}{x_{m,f,b,d} \cdot w_d \cdot e^{elec}_{m,f,eu,b}} - \dsum_{m \in M}E^{elec}_{m,f,eu}}{\dsum_{m \in M}E^{elec}_{m,f,eu}} \right| \leq \epsilon^B_{f,eu} \label{eq:mes}
    \\
    & \left| \frac{\dsum_{f \in F}\dsum_{m \in M}\dsum_{d \in D}\dsum_{b \in B_d}{x_{m,f,b,d} \cdot w_d \cdot e^{elec}_{m,f,eu,b}} - \hat{E}_{eu}}{\hat{E}_{eu}} - \frac{\dsum_{f \in F}\dsum_{m \in M} E^{elec}_{m,f,eu} - \dsum_{f \in F}\dsum_{m \in M} E^{base}_{m,f,eu}}{\dsum_{f \in F}\dsum_{m \in M} E^{base}_{m,f,eu}} \right| \leq \epsilon^C_{eu} \label{eq:fuels} 
    \\
    & \dsum_{f \in F} \dsum_{m \in M} x_{m,f,b} = \dsum_{m \in M} N_{m, b} \quad \forall b \in B  \label{eq:count} \\
    & \dsum_{f \in F} x_{m,f,b} \geq N_{m,b} \quad \forall b \in B,\; m \in M \setminus \{\text{'Ref'}\} \label{eq:non_ref} 
    \\
    & \dsum_{f \in F} x_{m,f,b} \leq N_{m,b} \quad \forall b \in B,\; m = \text{'Ref'} \label{eq:ref} \\
    & x_{m,f,b} = 0 \quad \forall  b \in B; (m,f) \in \mathcal{I} \label{eq:inf} 
\end{align}
}

The measure adoptions are represented by the binary variable $x_{m,f, b}$, which captures whether an electrification measure ($m$) is applied to fuel ($f$) of a building ($b_d$) in each feeder ($d$) of the representative set. To capture total energy consumption when a measure is adopted, this variable is scaled by the weight of the feeder where the building is located ($w_b$) and multiplied by the end-use consumption in the high-electrification scenario ($e^{elec}_{m,f,eu,b}$). 

The objective function (\ref{eq:obj}) minimizes a combination of three types of errors, with relative weights given by $\beta$, and defined in constraints (\ref{eq:bdg})--(\ref{eq:fuels}). The errors are organized from the least important to the most important for the disaggregation objective, which implies that the $\beta$ weights should be selected such that $\beta_A << \beta_B << \beta_C$. The first (\ref{eq:bdg}) and second (\ref{eq:mes}) constraints capture the error at the electrification measure and fuel levels, respectively, and can be used when such information is available in the county-level electrification projection. Constraint (\ref{eq:bdg}) ensures that the measures implemented across all buildings match the county-level electricity consumption projected for each measure ($E^{elec}_{m,f,eu}$). The second error term (\ref{eq:mes}) is formulated in a similar but less restrictive manner, capturing differences between the feeder set and the projected electricity consumption aggregated across all measures applicable to each fuel.

The third error term (\ref{eq:fuels}) is the most important, as it captures the accuracy of the disaggregation relative to each end-use in the county-level electrification projection. Rather than focusing on absolute differences, as in the two previous error terms, it ensures that the relative variation in energy consumption across the buildings in the feeder set matches the county-level variation in the energy consumption of the electrification scenario relative to that of the base year. This is necessary because, although the feeder sets are constructed to match base-year consumption as closely as possible, differences in end-use consumption may still exist between the representative feeders and the county-level baseline. As explained in \cite{seres2025}, this occurs because the feeder selection process is designed to simultaneously reproduce multiple characteristics of the county, including annual energy consumption, peak demand behavior, and historical investment patterns. Therefore, the error term here emphasizes reproducing the electrification trajectory of each end-use rather than its absolute consumption. Absolute differences relative to the final county-level electricity consumption are addressed later in the cost calculation.

The remaining constraints bound the measure adoption variables. Constraints (\ref{eq:count}) ensure consistency in total building counts across electrification measures. Constraints (\ref{eq:non_ref}) allow the number of buildings adopting electrification measures to increase when the selected measure is not the reference one (`Ref`). When the measure is `Ref`, the number of buildings can only decrease or remain constant, as expressed in constraints (\ref{eq:ref}). Finally, constraints (\ref{eq:inf}) enforce that measure adoption is zero when combinations of measures and fuels are not compatible, as defined by the infeasibility set $\mathcal{I}$.

\subsection{Feeder-level needs assessment}

The distribution upgrade assessment is performed through an iterative power-flow-based distribution planning algorithm (described in \ref{app:ugprade_costs}), implemented in LBNL's  LODGElite, a version of the LBNL Least-cost Optimal Distribution Grid Expansion (LODGE) model. For each representative feeder, the analysis is conducted at the hour corresponding to the annual feeder peak demand under the electrification scenario. A three-phase power flow is first executed to evaluate thermal loading and voltage violations relative to ANSI voltage limits. Whenever a line or transformer exceeds its thermal rating, its capacity is increased to the next available size in the corresponding equipment catalog. Similarly, voltage violations are addressed through the addition or resizing of voltage support devices. Capacitor banks are installed and, if violations persist, their capacities are increased in subsequent iterations by selecting the next available capacitor size. This iterative process continues until all thermal and voltage constraints are satisfied.

Voltage support devices are placed using a graph-based clustering approach to mimic voltage control zones used in real distribution system operation. Buses with voltage violations are first grouped according to the magnitude and temporal pattern of their violations using K-means clustering. Each cluster is then partitioned into electrically coherent voltage control zones using spectral clustering applied to the feeder topology. For zones with only undervoltage conditions, a capacitor bank is placed at the node farthest from the substation. For zones with overvoltages, a regulator is installed at the node closest to the substation. These locations are preserved throughout the iterative process, while the size of the installed capacitors may be increased as needed to eliminate remaining violations.

Following each upgrade iteration, a new three-phase power flow is executed and all thermal and voltage constraints are re-evaluated. When no additional violations exist in the three-phase part of the circuit, the final distribution upgrade plan is obtained by recording the final capacity selected for each line, transformer, and capacitor location throughout the iterative process. These final equipment ratings are then used to compute the feeder-level distribution upgrade costs.

\subsection{County-level distribution costs}

The distribution upgrades for each feeder obtained from the needs assessment are transformed into costs using the equipment unit costs presented in \ref{app:unit_costs}. Let $C_d^{elec}$ denote the upgrade cost obtained for representative distribution feeder $d$ under the electrification scenario, and let $w_d$ denote the weight (or number) of feeders of that type in the county. The total modeled county upgrade cost is then calculated as

\begin{align}
C_{\mathrm{modeled}}^{elec}
=
\sum_{d \in \mathcal{D}}
w_d \cdot
C_d^{elec}
\end{align}

where $\mathcal{D}$ is the set of representative distribution feeders.

As discussed above, although the representative feeder set is designed to reproduce the aggregate characteristics of the county distribution system, it is not expected to exactly match the baseline energy consumption. In addition, when disaggregating the electrification projections, the most relevant objective term, defined in equation (\ref{eq:fuels}), puts more emphasis on reproducing the electrification changes in end-use demand relative to base year than on correcting absolute differences in energy consumption. Therefore, differences between the energy represented by the feeder set and the aggregated county baseline may persist in the electrification scenario.

Thus, the total energy consumption represented by the disaggregated feeder simulations, denoted by $E_{\mathrm{modeled}}^{elec}$, may differ from the corresponding aggregated county-level projection, denoted by $E_{\mathrm{actual}}^{elec}$. To account for this difference when estimating county-wide distribution upgrade costs, the modeled cost is normalized by the energy represented by the feeder set. The resulting cost per unit of energy is calculated as:

\begin{align}
E_{\mathrm{modeled}}^{elec}
&= \dsum_{eu \in EU}\dsum_{f \in F}\dsum_{m \in M}\dsum_{d \in D}\dsum_{b \in B_d}x_{m,f,b,d}
\\
c^{elec}
&= \frac{C_{\mathrm{modeled}}^{elec}}{E_{\mathrm{modeled}}^{elec}}
\end{align}

The adjusted county-wide upgrade cost is then obtained by applying this value to the aggregated county-level energy projection

\begin{align}
E_{\mathrm{actual}}^{elec}
&= \dsum_{eu \in EU}\dsum_{f \in F}\dsum_{m \in M} E^{elec}_{m,f,eu} \\
C_{\mathrm{actual}}^{elec}
& =
c^{elec} \cdot E_{\mathrm{actual}}^{elec}
\end{align}

where $C_{\mathrm{actual}}^{elec}$ represents the final estimate of distribution upgrade costs associated with the county-level electrification scenario.

\section{Case Study}

To demonstrate the methodology described above, we use the case of Genesee County, located in east-central Michigan, approximately 65 miles northwest of Detroit and centered around the city of Flint. The county has a population of approximately 400,000 people, making it one of the largest counties in Michigan. It is primarily suburban, with an urban center in Flint and rural areas along the county’s outer edges. 

Input and output data of this case study can be found in \cite{Heleno2026GeneseeResults}, where aggregated baseline and high-electrification projections are obtained from \cite{Langevin2026}. The baseline was considered the year of 2024, where the total electricity consumption of the commercial and residential building stock in the county is 2.65~TWh, with a peak demand of 615~MW. In this analysis, we consider the high-electrification scenario for 2035,where the total building electricity consumption increases to 2.78~TWh.

\subsection{County representative feeder sets}

By applying the methodology described in \cite{seres2025}, we obtain the representative set of feeders summarized in Table \ref{tab:feeders}. The results include a set of feeder types with associated weights that express the number of times each representative feeder must be replicated to match the county-wide electricity consumption and distribution infrastructure costs. The resulting set includes one large-scale high-density urban feeder (feeder \#1), one exclusively commercial feeder (feeder \#3), and seven primarily residential feeders (feeders \#2 and 4-9) with smaller shares of commercial buildings. For more details on the specific characteristics of the taxonomy suburban feeders, please refer to \cite{Schneider2008ModernReport}. 
 
\begin{table}[ht]
\centering
\begin{tabular}{c c l p{2.5cm}p{2.2cm} p{2.2cm}}
\hline
\textbf{Feeder} &
\textbf{weight} &
\textbf{Type} &
\textbf{Taxonomy feeder version} &
\centering\textbf{Residential bdg \# } &
\centering\textbf{Commercial bdg \# } \tabularnewline
\hline
1 & 1  & IEEELVTestCaseNorthAmerican & 51 & 2748 & 226 \\
2 & 1  & R2\_12\_47\_1              & 6  & 492  & 34  \\
3 & 1  & GC\_12\_47\_1              & 42 & 0    & 19  \\
4 & 6  & R2\_12\_47\_3              & 14 & 2714 & 11  \\
5 & 1  & R2\_25\_00\_1              & 17 & 2346 & 60  \\
6 & 1  & R2\_35\_00\_1              & 8  & 526  & 4   \\
7 & 1  & R2\_35\_00\_1              & 36 & 589  & 6   \\
8 & 19 & R2\_12\_47\_2              & 15 & 2123 & 20  \\
9 & 61 & R2\_12\_47\_2              & 16 & 2293 & 9   \\
\hline
\end{tabular}
\caption{Selected feeders and associated building counts.}
\label{tab:feeders}
\end{table}

\subsection{Disaggregation of building electrification projection}

\begin{figure}[ht]
    \centering
    \includegraphics[width=\linewidth]{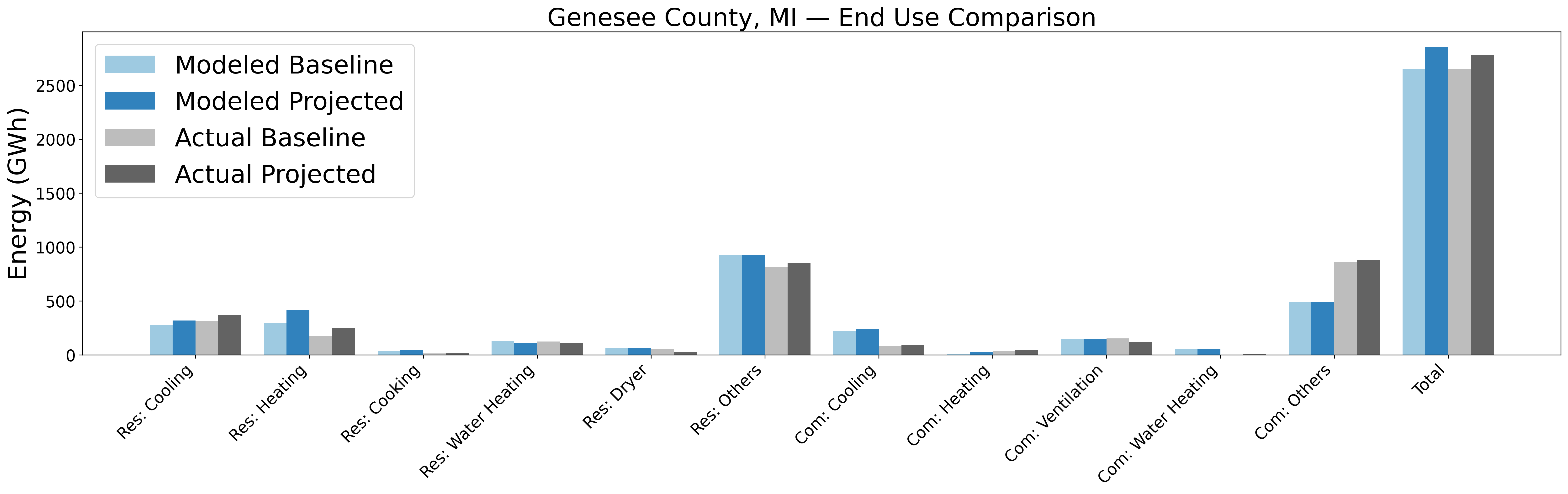}
    \caption{Comparison of end-use electricity consumption profiles for the representative building stock in Genesee County. The figure presents, for each end-use, county-level "actual" and "modeled" energy-consumption projections. "Baseline" refers to year 2024 and "Projected" refers to the high-electrification scenario of the year 2035.}
    \label{fig:end_use_comparison}
\end{figure}

The model proposed in this work to disaggregate county-level projections of high-electrification, and described in equations (\ref{eq:obj})-(\ref{eq:inf}), was applied to the end-use projections for Genesee County. Figure \ref{fig:end_use_comparison} compares the aggregated feeder-level projections ("modeled") with the "actual" county-level energy projections from the 2035 high-electrification scenario. The figure also includes an analogous comparison for the baseline scenario resulting from the feeder generation methodology. By comparing the "actual baseline" and "actual projected" bars (both depicted in grey), it is possible to observe an increase in total electricity consumption in Genesee County from 2024 to 2035, arising primarily from increased residential heating and cooling electricity demand. The aggregated "modeled results", obtained from the optimization model (\ref{eq:obj})--(\ref{eq:inf}), reproduce similar variations in end-use electricity consumption to those observed in the "actual" projections. It is important to note that the focus here is primarily on reproducing the variation in energy consumption rather than minimizing absolute differences. As discussed above, the third error term in the objective function, given by (\ref{eq:fuels}), is formulated to ensure that the relative variation in energy consumption for each end-use in the feeder set matches the projected variation in the high-electrification scenario relative to the 2024 base year. The absolute difference between "modeled" and "actual" projected consumptions will be taken into account when calculating total costs.

To better illustrate the results of the end-use level optimization (\ref{eq:obj})-(\ref{eq:inf}), we track the decision variables associated with the adoption of electrification measures for the HVAC, with impact in heating and cooling end-uses. The electrification of space heating and cooling end-uses can be achieved through the adoption of different types of heat pump technologies, combined with fuel switching and building envelope enhancements. Table \ref{tab:residential_hvac_measures} presents the 9 HVAC electrification measures available to residential buildings in Genesee County.

\begin{table}[ht]
\centering
\caption{Residential HVAC electrification measures considered in the analysis.}
\label{tab:residential_hvac_measures}
\begin{tabular}{ll}
\hline
\textbf{HVAC Measure} & \textbf{Label} \\
\hline
ENERGY STAR HP -- Partial Fuel Switching &
(R) ESTAR HP Pt. FS \\

ENERGY STAR HP -- Full Fuel Switching &
(R) ESTAR HP TS \\

High-Efficiency HP -- Full Fuel Switching &
(R) Best HP TS \\

ENERGY STAR HP + Envelope -- Partial Fuel Switching &
(R) ES HP Pt. FS \& Env. \\

High-Efficiency HP -- Partial Fuel Switching &
(R) Best HP Pt. FS \\

ENERGY STAR HP + Envelope -- Full Fuel Switching &
(R) ES HP TS \& Env. \\

High-Efficiency HP + Enhanced Envelope -- Full Fuel Switching &
(R) Bst. HP TS \& Env.+ \\

High-Efficiency HP -- Fuel Switching &
(R) Best HP FS \\

High-Efficiency HP + Enhanced Envelope -- Fuel Switching &
(R) Bst. HP FS \& Env.+ \\
\hline
\end{tabular}
\end{table}

\begin{figure}[ht]
    \centering
    \includegraphics[width=\linewidth]{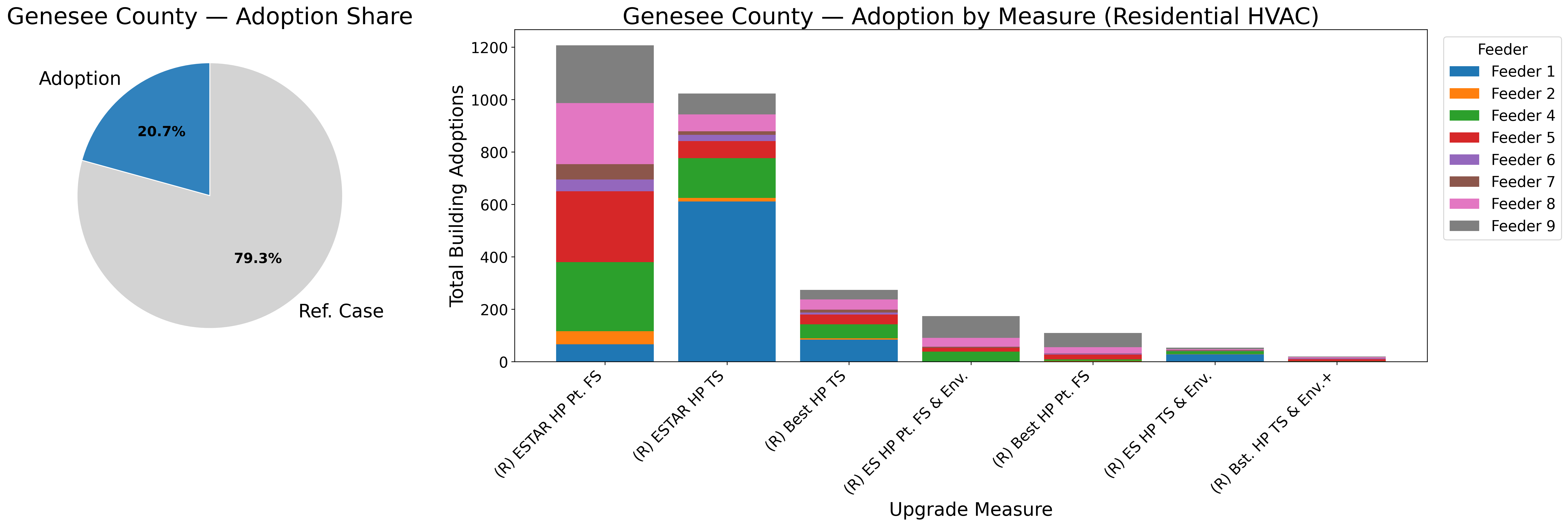}
    \caption{Distribution of residential electrification measures for HVAC end-uses in Genesee County. Left: Share of buildings that adopted an electrification measure. Right: distribution of adopted measures per feeder set.}
    \label{fig:county_adoption_summary}
\end{figure}

Figure \ref{fig:county_adoption_summary} summarizes the adoption of residential HVAC measures resulting from model (\ref{eq:obj})--(\ref{eq:inf}). The pie chart in the left-hand panel provides an overview of HVAC measure adoption across all residential consumers in the county. It can be seen that, to achieve the prescribed county-level projections of heating and cooling end-uses, 20.7\% of residential consumers adopted some form of HVAC electrification. The right-hand panel shows how the adopted electrification measures are distributed across the representative feeders. The building adoption numbers refer to adoptions in the individual representative feeders, not taking into account the feeder weights. As expected, most residential HVAC electrification occurs in feeders 1, 4, 5, 8, and 9, which serve the largest number of residential consumers. The standard ENERGY STAR air-sourced heat pump, with either partial (heating or cooling only) or complete fuel switching, is the most widely adopted technology.

\begin{figure}[ht]
    \centering
    \includegraphics[width=0.8\linewidth]{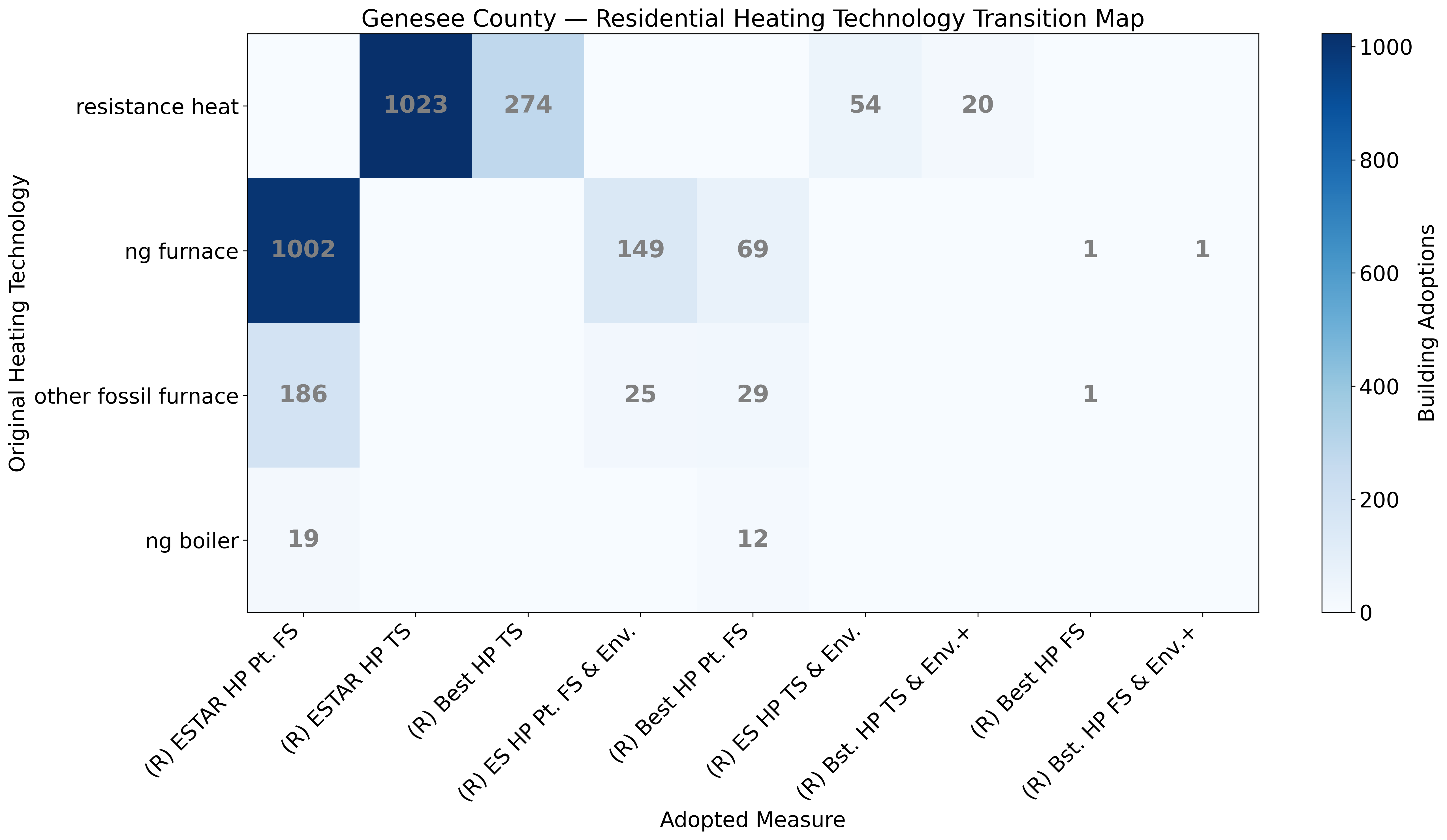}
    \caption{HVAC measure adoption transition to match county-level residential heating projections. Numbers refer to the number of buildings in the representative feeder set.}
    \label{fig:heating_transition}
\end{figure}

The chart in Figure \ref{fig:heating_transition} presents a different perspective on the same decision variables of model (\ref{eq:obj})-(\ref{eq:inf}) by illustrating the technology transitions that enabled residential space-heating electrification. It can be seen that a portion of the county's residential heating electrification is achieved through full or partial fuel switching, primarily from natural gas systems, while the remaining portion results from the replacement of existing electric resistance heating systems with more efficient air-sourced heat pumps. Consequently, some heating electrification measures increase electricity consumption by shifting heating demand from fossil fuels to electricity, whereas others decrease electricity consumption by improving the efficiency of already electrified heating systems. The combined effect is still a net increase in residential electricity consumption for space heating, consistent with the county-level electrification projections, as shown in figure \ref{fig:end_use_comparison}.

An interesting aspect of the optimization results is that the model does not rely exclusively on fuel switching to achieve the required increase in heating electrification. Instead, it also selects the replacement of electric resistance heating systems with heat pumps, even though this reduces electricity consumption for heating. This behavior can be explain by the fact that the HVAC measures considered are all air-source heat pumps that simultaneously affect both heating and cooling end-uses. This means that, in many cases, the model adopts heat pumps primarily to satisfy the projected increase in residential cooling electrification, while the associated reduction in heating electricity consumption is offset by additional fuel-switching measures elsewhere in the building stock.

\subsection{Feeder upgrade needs and county distribution grid costs}

\begin{figure}[ht]
    \centering
    \includegraphics[width=\linewidth]{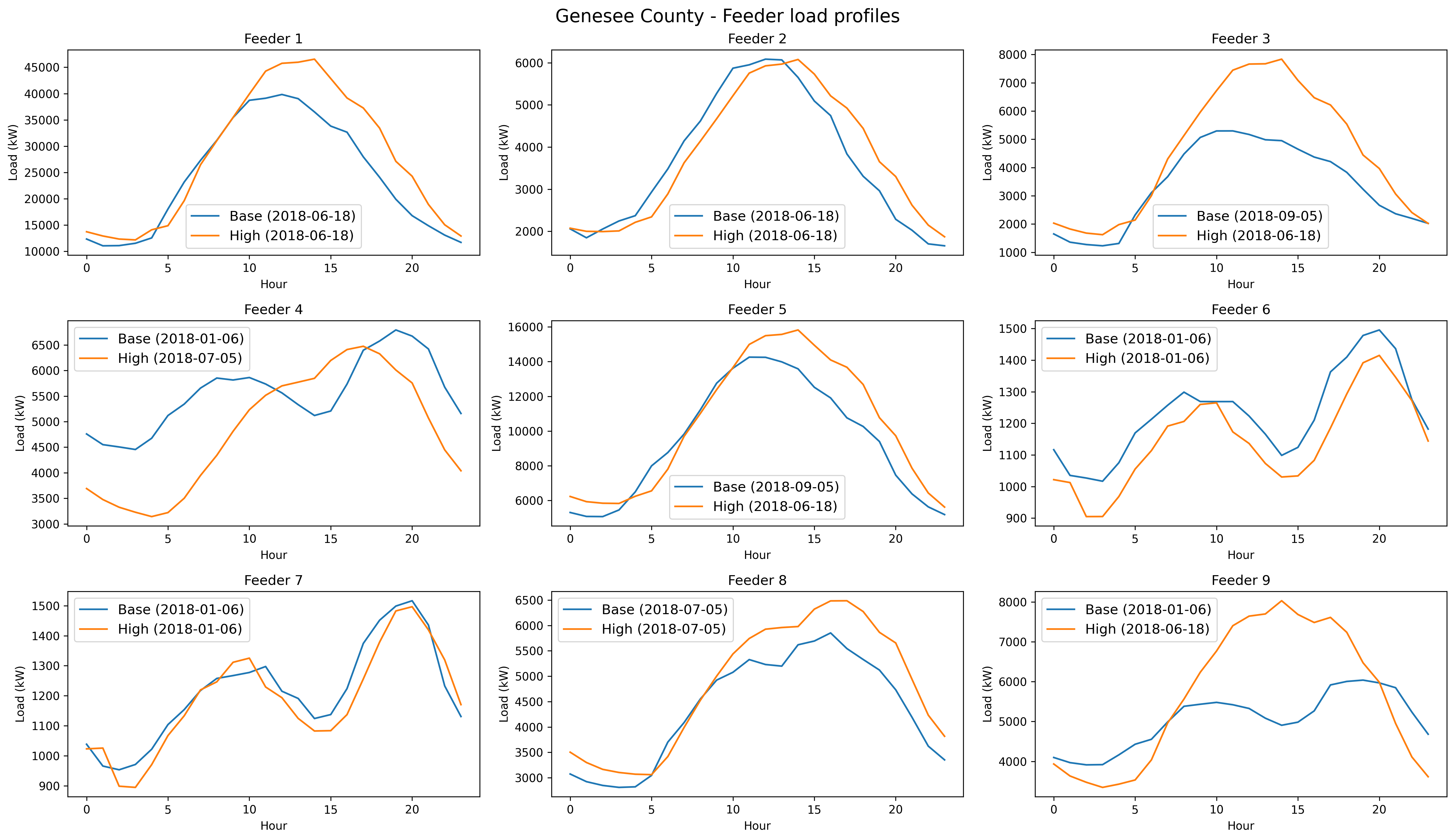}
    \caption{Comparison of building load profiles per feeder in the representative feeder set on the day of the feeder annual peak demand, between the baseline ("Base") and high-electrification ("High") scenarios. The legend indicates the date on which the peak profile occurs.}
    \label{fig:load_profiles}
\end{figure}

After disaggregating the electrification measures at the building level, it is possible to capture hourly variations in building electricity demand and reconstruct the new load profiles at the feeder level. Figure \ref{fig:load_profiles} presents the load profiles on the day of the feeder coincident peak for the 9 representative feeders in Genesee County, considering both the baseline 2024 scenario ("base") and the high-electrification 2035 scenario ("High").

The results show that building electrification measures affect feeder-level load profiles in different ways. In feeders 6 and 7, the annual peak load in the baseline scenario occurs in January, driven by residential electric resistance heating demand. In the high-electrification scenario, the peak occurs on the same day but is slightly reduced, indicating the replacement of resistance heating with more efficient electric heating technologies. Feeders 4 and 9 also have winter peaks in the baseline scenario, but their peaks shift to the summer under high-electrification due to increased cooling electricity demand. In feeder 4, however, the summer peak becomes dominant mainly because of a reduction in winter heating demand rather than a large increase in summer cooling load, resulting in an overall decrease in peak demand. Feeders 1, 2, and 8 maintain the summer peaks observe in the baseline scenario, but the magnitude of their peaks increase due to new cooling electric loads, with the exception of feeder 2 that remains constant. A similar effect is observed in feeders 3 and 5, where peak demand shifts from late summer to early summer as cooling demand increases.

We mapped the resulting disaggregated electrification measures at the building level to the corresponding nodes of the representative distribution feeders in Genesee County and evaluated the need for distribution system upgrades in each feeder. Table \ref{tab:upgrade_costs} presents the total upgrade costs obtained for each feeder, together with the corresponding breakdown of investments by upgrade category, including lines, transformers, and capacitors. Voltage regulators were not selected as an investment option because this work only considers electrification measures that affect peak demand (i.e., the electrification measures do not include distributed generation). Consequently, the identified upgrade needs are primarily driven by capacity constraints and low-voltage conditions, which can be addressed by reconductoring, transformer upgrades, and capacitors.

The electrification measures resulted in a total modeled distribution upgrade cost of 52.23 M\$ (2024 dollar value) based on the representative feeder set. The largest share of these costs is associated with service transformer upgrades, which are primarily driven by localized increases in load, followed by distribution line reconductoring. Capacitor investments account for approximately 10\% of the total upgrade costs.

\begin{table*}[ht]
\centering
\caption{Distribution upgrade costs and feeder peak loads under the 2035 high-electrification scenario.}
\label{tab:upgrade_costs}
\small
\begin{tabular}{lccccccc}
\hline
Feeder & Weight & Total & Lines & Transformers & Capacitors & Baseline Peak & 2035 Peak \\
& & (k\$) & (k\$) & (k\$) & (k\$) & (MW) & (MW) \\
\hline
Feeder 1 & 1  & 224.2  & 224.2  & 0.0   & 0.0   & 39.83 & 46.54 \\
Feeder 2 & 1  & 0.0    & 0.0    & 0.0   & 0.0   & 6.08  & 6.08 \\
Feeder 3 & 1  & 405.9  & 12.9   & 393.0 & 0.0   & 5.30  & 7.84 \\
Feeder 4 & 6  & 327.8  & 130.8  & 94.6  & 102.4 & 6.79  & 6.47 \\
Feeder 5 & 1  & 3314.6 & 2838.5 & 396.3 & 79.8  & 14.26 & 15.82 \\
Feeder 6 & 1  & 0.0    & 0.0    & 0.0   & 0.0   & 1.50  & 1.41 \\
Feeder 7 & 1  & 8253.4 & 8172.7 & 5.0   & 75.7  & 1.52  & 1.50 \\
Feeder 8 & 19 & 78.3   & 0.0    & 78.3  & 0.0   & 5.85  & 6.49 \\
Feeder 9 & 61 & 599.7  & 0.0    & 515.3 & 84.4  & 6.04  & 8.03 \\
\hline
\textbf{County Total} &
\textbf{} &
\textbf{52.23 M\$} &
\textbf{12.03 M\$} &
\textbf{34.28 M\$} &
\textbf{5.92 M\$} &
&
\\
\hline
\end{tabular}
\end{table*}

It is interesting to note that feeders \#8 and \#9, the two most representative feeders according to Table~\ref{tab:feeders}, experience significant increases in coincident peak demand, yet they require substantially different transformer investments. This occurs because transformer upgrades are driven by localized increases in demand at individual nodes rather than by the aggregate feeder peak demand. This underscores the need to capture these localized variations through modeling the spatial distribution of the electrification measures.

\begin{figure}[ht]
    \centering
    \includegraphics[width=\linewidth]{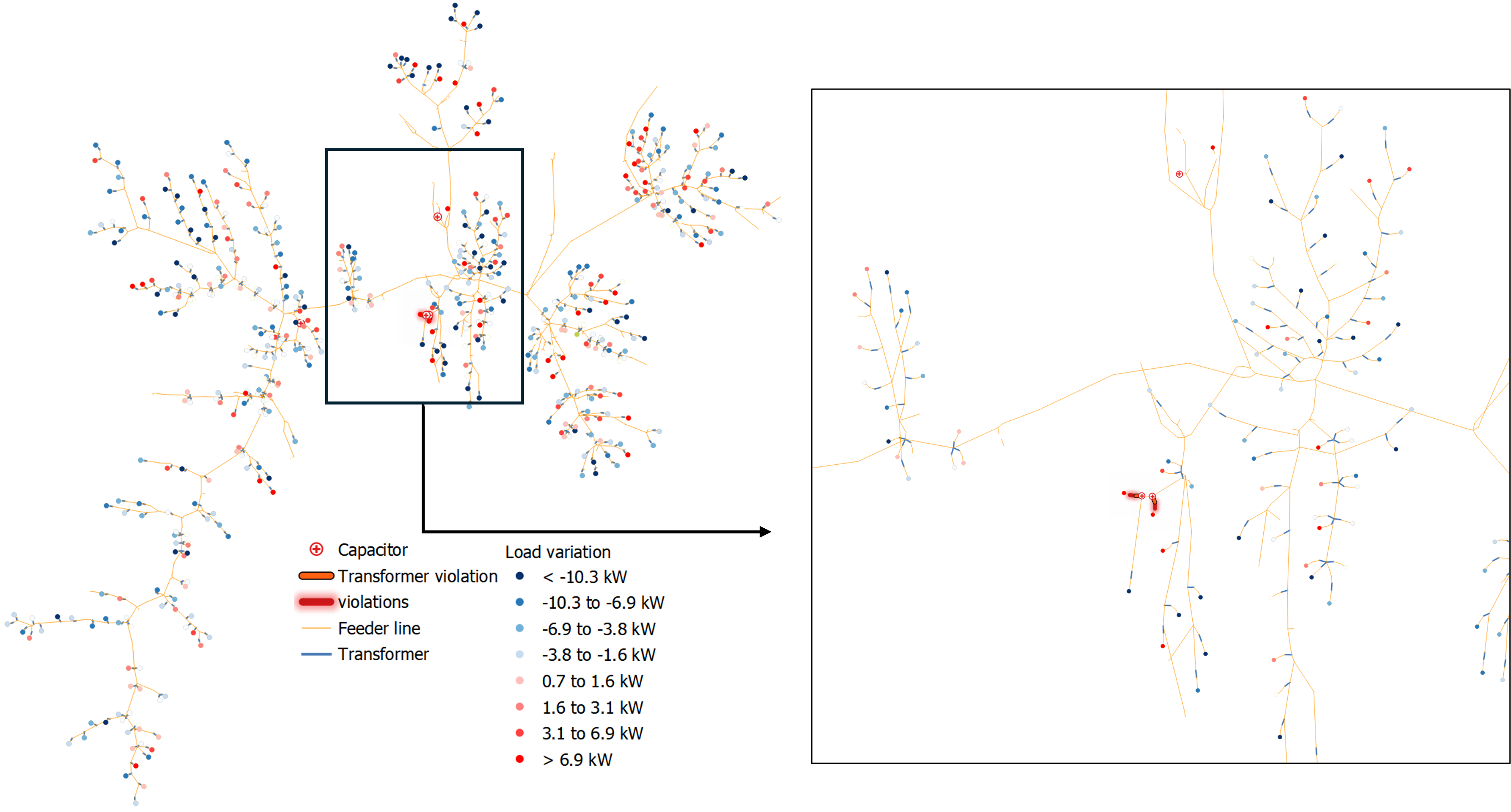}
    \caption{Feeder 4 (R2\_12\_47\_1) with investments required in a specific location. Colors indicate the variation in nodal loads between the electrification and the baseline scenario.}
    \label{fig:feeder4_upgrades}
\end{figure}

Another interesting aspect worth highlighting is that the third most representative feeder (feeder 4) requires line and capacitor upgrades even though its total peak demand decreases in the high-electrification scenario. Once again, this result is explained by the spatial distribution of the electrification measures. To illustrate this effect, Figure \ref{fig:feeder4_upgrades} shows the changes in nodal loads across feeder 4 together with the resulting network investments. It can be seen that, although the electrification measures reduce the feeder's overall peak demand, there are specific areas where load increases. These localized increases create voltage and capacity constraints that trigger line and capacitor upgrades. As a result, infrastructure investments become concentrated in particular sections of the feeder, despite the reduction in aggregate peak demand.

As discussed in the methodology section, the total energy consumption represented by the feeder set does not exactly match the county-level load projections. To obtain a county-wide cost estimate, the modeled upgrade costs are adjusted using the ratio between the actual projected county energy consumption and the energy consumption represented by the feeder set. The modeled feeder set represents 2,853 GWh of annual electricity consumption, while the corresponding county-level projection is 2,782 GWh. The adjustment is performed by first calculating the modeled upgrade cost per unit of energy consumption and then applying this value to the projected county load. The resulting adjusted distribution upgrade cost is 50.94 M\$.

\begin{align}
c^{\mathrm{adj}}
&=
\frac{52.23 \times 10^6}{2{,}853 \times 10^6}
=
0.01831 \ \$/\mathrm{kWh}
\end{align}

\begin{align}
C^{\mathrm{actual}}
&=
\frac{0.01831 \times 2{,}782 \times 10^6}{10^6}
=
50.94 \ \mathrm{M\$}
\end{align}

\section{Conclusion}

This paper presented a methodology to disaggregate county-level building electrification projections onto representative distribution feeders and evaluate their impact on distribution upgrade costs. The results show that the optimization model can accurately reproduce the projected changes in aggregate end-use electricity consumption by assigning detailed electrification measures to individual buildings across the representative feeder set.

The case study of Genesee County demonstrates that the methodology is able to capture a diversity of feeder-level responses to electrification and represent the different realities that building electrification produces within the distribution system. Some feeders experienced increases in peak demand, while others experienced reductions. In some feeders, peak demand shifted from winter to summer, while in others the peak season remained unchanged. These differences emerged from the composition of the building stock, the baseline technologies, and the allocation of electrification measures across the feeder set.

The case study also illustrates the value of representing electrification at the building level when estimating distribution infrastructure costs. Localized changes in demand created upgrade needs that were not directly related to changes in aggregate feeder peak demand. In some cases, infrastructure investments were required even when feeder-level peak demand decreased. These results show that preserving the spatial distribution of electrification measures provides information that cannot be captured through aggregate energy projections or marginal distribution cost assumptions.

Future work can extend the methodology beyond building electrification to include other drivers of electricity demand that affect the distribution system, such as electric vehicle adoption, distributed generation, energy storage, and demand response. Unlike building electrification, these technologies can introduce changes that extend beyond peak demand, for example including reverse power flow. Capturing these effects may require extending the current framework to include multi-temporal distribution system simulations, allowing the assessment of infrastructure costs associated with a broader range distribution system load.

\section*{Acknowledgements}
This work was funded by the U.S. Department of Energy under the contract DE-AC02-05CH11231. The views expressed herein do not necessarily represent the views of the DOE or the U.S. Government.

\newpage
\appendix

\section{Distribution grid upgrade algorithm}
\label{app:ugprade_costs}

\begin{algorithm}[H]
\caption{Distribution Upgrade Assessment}
\label{alg:distribution_upgrade}
\begin{algorithmic}[1]
\Require Feeder $f$, peak-hour load profile $\mathcal{L}_f$, catalogs of lines, transformers and capacitors.
\Ensure Final equipment upgrades and total upgrade cost $U^{f}_{cost}$.

\State Initialize existing equipment ratings and capacitor locations.
\State $finish \gets$ False

\While{$finish =$ False}

    \State Run three-phase power flow.
    \State Obtain overloaded lines $\mathcal{L}^{viol}$, overloaded transformers $\mathcal{T}^{viol}$, and voltage violations $\mathcal{B}^{viol}$.
    \State $finish \gets$ True

    \For{$l \in \mathcal{L}^{viol}$}
        \State Upgrade $l$ to the next available conductor size.
        \State $finish \gets$ False
    \EndFor

    \For{$t \in \mathcal{T}^{viol}$}
        \State Upgrade $t$ to the next available transformer size.
        \State $finish \gets$ False
    \EndFor

    \If{$\mathcal{B}^{viol} \neq \emptyset$}

        \State Construct voltage violation matrix $V_{b,t}$.
        \State Obtain voltage-control zones using KMeans Bus Voltage Classification followed by Spectral Clustering Zone Division.

        \For{each zone $\mathbb{Z}$}

            \If{all violations in $\mathbb{Z}$ are undervoltages}

                \State Select the bus farthest from the substation.
                \State Install a capacitor or increase to the next capacitor size.

            \Else

                \State Select the bus closest to the substation.
                \State Install a voltage regulator.

            \EndIf

        \EndFor

        \State $finish \gets$ False

    \EndIf

\EndWhile

\State Record the largest line, transformer, and capacitor size selected at each location.
\State Compute total upgrade cost $U^{f}_{cost}$ from the final equipment selections.

\State \Return $U^{f}_{cost}$ and final equipment upgrades.

\end{algorithmic}
\end{algorithm}

\section{Distribution grid unit costs}
\label{app:unit_costs}

Distribution system upgrade costs were taken from NLR's Distribution Grid Integration Unit Cost Database \cite{cost_db}, which compiles utility-based cost estimates for distribution system components and was specifically developed to support distribution grid integration studies. Transformer and capacitor bank costs were represented using the discrete upgrade options presented in Tables~\ref{tab:trafo_costs} and \ref{tab:capacitor_costs}, respectively.

For line reconductoring, the set of conductor upgrade options presented in Table~\ref{tab:line_costs} was considered. Each option is characterized by its installation cost, ampacity, and electrical parameters. The resistance and reactance values associated with each conductor were obtained from \cite{Kersting_book}. These component costs and technical characteristics were used to estimate the distribution system investments required to accommodate the projected electrification scenarios.

In the tables below, most values are reported in 2016 dollars and were multiplied by 1.31 to approximate 2024 dollar values in the results.

\begin{table}[ht]
\centering
\caption{Distribution line upgrade options used in the distribution network model. Costs correspond to non-California installations and are reported in thousands of dollars per mile (k\$/mile).}
\label{tab:line_costs}
\begin{tabular}{lccccc}
\hline
Size & Type & Capacity & Cost & $R$ & $X$ \\
& & (A) & (k\$/mile) & ($\Omega$/mile) & ($\Omega$/mile) \\
\hline
\#4    & OH & 140 & 644.2  & 2.57 & 0.25 \\
\#4    & UG & 140 & 174.2  & 2.57 & 0.25 \\
\#2    & OH & 180 & 818.1  & 2.07 & 0.24 \\
\#2    & UG & 180 & 221.3  & 2.07 & 0.24 \\
1/0    & OH & 230 & 1230.3 & 1.12 & 0.23 \\
1/0    & UG & 230 & 332.8  & 1.12 & 0.23 \\
3/0    & OH & 300 & 1874.5 & 0.72 & 0.22 \\
3/0    & UG & 300 & 507.0  & 0.72 & 0.22 \\
4/0    & OH & 340 & 2344.7 & 0.59 & 0.22 \\
4/0    & UG & 340 & 634.2  & 0.59 & 0.22 \\
336.4  & OH & 530 & 3633.1 & 0.31 & 0.21 \\
336.4  & UG & 530 & 982.7  & 0.31 & 0.21 \\
477    & OH & 670 & 5037.3 & 0.22 & 0.21 \\
477    & UG & 670 & 1362.6 & 0.22 & 0.21 \\
\hline
\end{tabular}
\end{table}

\begin{table}[ht]
\centering
\caption{Transformer upgrade costs used in the distribution network model.}
\label{tab:trafo_costs}
\begin{tabular}{cc}
\hline
Capacity (kVA) & Cost (k\$) \\
\hline
25    & 3.85 \\
50    & 4.18 \\
75    & 5.25 \\
100   & 6.06 \\
112.5 & 8.34 \\
150   & 12.50 \\
225   & 13.91 \\
300   & 35.00 \\
500   & 44.00 \\
750   & 51.00 \\
1000  & 66.00 \\
1500  & 92.00 \\
2500  & 146.00 \\
3500  & 250.00 \\
5000  & 300.00 \\
6000  & 380.00 \\
10000 & 600.00 \\
12000 & 800.00 \\
\hline
\end{tabular}
\end{table}

\begin{table}[ht]
\centering
\caption{Capacitor bank upgrade options used in the distribution network model.}
\label{tab:capacitor_costs}
\begin{tabular}{ccc}
\hline
Capacity & Switched & Cost \\
(kVar) &  & (k\$) \\
\hline
100  & No  & 7.0 \\
300  & No  & 9.0 \\
600  & Yes & 10.7 \\
900  & Yes & 13.7 \\
1200 & Yes & 32.2 \\
\hline
\end{tabular}
\end{table}

\clearpage
\bibliography{references}

\end{document}